\documentclass[aps,final,notitlepage,oneside,twocolumn,nobibnotes,nofootinbib,
superscriptaddress,noshowpacs,centertags]{revtex4-1}

\usepackage[utf8]{inputenc}
\usepackage[english]{babel}
\usepackage{graphicx}
\usepackage{latexsym}
\usepackage{amssymb}
\usepackage{amsmath}
\usepackage{float}

\begin{document}

\title{Activity of the ejected galactic nuclei and cosmic rays}
\author{Yury Eroshenko}\thanks{e-mail: eroshenko@inr.ac.ru}
\affiliation{Institute for Nuclear Research of the Russian Academy of Sciences, 60th October Anniversary Prospect 7a, Moscow, 117312, Russia}

\date{\today}

\begin{abstract}
The double and triple approaches of supermassive central black holes are a fairly common process in the hierarchical picture of galaxy formation. In this case, the ejection of a black hole is possible; it may carry the accretion disk and some amount of stars with it. The activity of the galactic nucleus can be restored after the ejection. One consequence of this activity is an ultrafast gas outflow, accompanied by a shock wave. We show that the acceleration of cosmic rays at such a shock is more effective than in the case of a stationary galactic nucleus due to the lower density of the surrounding gas and the higher velocity of the shock front. In this case, proton acceleration can occur up to energies of $\sim10^{19}$~eV, which is an order of magnitude higher than in an ordinary nucleus in the centers of galaxies.
\end{abstract}

\maketitle


\section{Introduction}

The origin of ultrahigh-energy cosmic rays (UHECRs) is one of the main problems of astrophysics and high-energy physics, see \cite{Ber13-1,Ber13-2,NorMelAch95}. Models of UHECR generation in jets from active galactic nuclei (AGNs) at shock waves in a magnetic field and at the boundaries of jets have been proposed  \cite{Ost90,Ost98,Ost00}. Various aspects of UHECR acceleration are considered in \cite{NorMelAch95,RacBie93,Goretal04,McKBla09}. One of the main factors limiting the maximum achievable energy during acceleration is the synchrotron losses of particles in magnetic fields \cite{BieStr87}. Comparable losses occur due to the scattering of particles by the radiation field of stars in the galaxy \cite{NorMelAch95,BieStr87}. In the case of a short jet that does not go beyond a galaxy, this problem is most severe due to the high energy density of radiation inside galaxies \cite{Ber77}.

In the work \cite{WanLoe17}, a model of acceleration of UHECRs in the ultrafast gas outflows (UFOs) from AGN was proposed. Such UFOs with velocities of $(0.1-0.4)c$ at distances of $10^2-10^4$ gravitational radii from the central supermassive black hole (SMBH) are observed in about half of the active galaxies \cite{Tometal10}. In this model, the UHECRs are accelerated in the region of the shock wavethat is formed near the front of the UFO. 

In \cite{WanLoe17}, the most optimistic parameters of the model were used, and it was concluded that it is possible to accelerate the UHECRs to extremely high energies $\sim10^{20}$~eV (100~EeV). In \cite{Peretal23,EhlOikPer25}, numerical calculations were performed under more conservative assumptions, and it was shown that only an energy of $\sim0.1$~EeV can be achieved, that in 5-10~\% of the galaxies UHECR acceleration occurs up to $\sim10^{0.4}$~EeV, and only in some rare cases does the energy of accelerated particles reach $\sim10^{1.8}$~EeV. According to these results, the maximum energy of the UHECRs is limited by their interaction with the radiation field of the disk, the corona of the disk, and the massive dusty torus present in the center of the galaxy.  

In this paper, we show that the above-mentioned problem of radiation field losses is significantly weakened if the generation of UHECRs occurs not in the central core of the galaxy, but in a core that has flown out of the central region due to the effect of gravitational wave emission or slingshot ejection during the double or triple gravitational interactions of SMBHs, respectively. According to modern concepts, galaxy mergers are a typical stage of their evolution. Galaxies are formed hierarchically by merging smaller protogalaxies. About half of the observed galaxies are in the process of mergers or interactions. Merging galaxies are most often seen in galaxy clusters, where the concentration of galaxies is very high. The nuclei of merging galaxies, under the influence of dynamic friction, move towards the common center of the new galaxy, where they unite. If the galaxies differ markedly in mass, then the smaller galaxy orbits inside the larger galaxy, loses mass due to tidal stripping, and eventually falls into the center of the larger galaxy. The central SMBHs of merging galaxies should also eventually merge. 

There are two possible mechanisms for ejecting a nucleus from the center of a galaxy. In the first mechanism (gravitational recoil), an SMBH formed as a result of the merger of two SMBHs is ejected due to the asymmetric emission of gravitational waves during the merger \cite{Fit83,ZloCamLou11,Donetal25}. The characteristic ejection velocity is $v_{\rm esc}\sim 100-300$~km~s$^{-1}$. The second mechanism is based on triple approaches of SMBHs and the effect of a slingshot ejection \cite{SasValAar74}. In this second mechanism, the ejection speed can be one to two orders of magnitude higher than in the first one. As shown in the works \cite{ReeSas75,ValBas91}, the ejected galactic nucleus can maintain its quasar activity due to interaction with the gas along its path.

Let us describe the second mechanism in more detail. In the hierarchical picture, a fairly typical case is the merger of not only two, but also almost simultaneously three or more galaxies. At the same time, it is possible for three nuclei of smaller galaxies to converge in the center of the combined galaxy at once. The most likely process is when two cores first approach each other, and later a third one approaches them. Since the final stage of the SMBH merger may be delayed (the problem of the last parsecs), the third SMBH may have time to settle to the center under the influence of dynamic friction.  At the same time, in the resulting triple system, the effect of a slingshot ejection is possible, which is known in the three-body problem \cite{Sas89}. In a slingshot, one of the bodies is ejected, and the other two move into a tight orbit. The ejection energy of the first body is derived from the change in the orbital energy of the other two bodies. More complex cases of triple interactions are also possible, such as exchange and ``ionization'' \cite{Sas89}. A numerical study of the process of slingshot ejection was performed in \cite{SasValAar74,LinSas77,MikVal90,HofLoe07}. In \cite{SasValAar74}, it was found that the probability of a triple approach and the ejection of an SMBH is $\sim10^{-4}-10^{-2}$. In this case, the number of ejected nuclei can be comparable to the number of active galactic nuclei.

The cores of galaxies are central black holes surrounded by an accretion disk and, possibly, a central star cluster and some other near-nuclear matter that was not stripped by tidal gravitational forces on the way to the center of the new galaxy. This matter collects in a dense gas-dust torus heated by radiation from a central source. It was shown in \cite{LinSas77} that an accretion disk around the ejected black hole is highly likely to be preserved during a slingshot ejection, and thus the ejected core can maintain its activity. An interesting question is the survival of the central part of the star cluster around the black hole. In the cores of galaxies, there are central star clusters \cite{Boe10}, the inner part of which has a stellar cusp around a SMBH. If the cusp or some part of it survives during the ejection, then in the future these surviving stars may become an additional source of gas supply for the accretion disk. 

In this paper, the interaction of accelerated particles with the radiation field of the disk around the ejected SMBH is calculated. In this case, approximately an order of magnitude higher maximum proton energies can be achieved. This is due to the faster movement of the shock wave in a medium with a reduced density, which leads to a rapid runaway of the wave from the area of strong radiation exposure of the disk. The presence of such UHECR sources inside the GZK horizon could ensure the appearance of UHECRs with almost maximum energies.


\section{Ejection of galactic nuclei}

In this Section, we discuss two mechanisms for ejecting a nucleus from the center of a galaxy during the approach of SMBHs. In the first mechanism, an SMBH formed as a result of the merger of two SMBHs is ejected due to the asymmetric emission of gravitational waves during the merger. The second mechanism is based on triple approaches of SMBHs and the effect of a slingshot ejection. In both cases, the nucleus may resume its activity through the accretion of surrounding gas.


  \subsection{Gravitational recoil ejection}
  \label{gwsec}
  
The recoil effect during the merger of two black holes has been studied in numerous works, see e.g. \cite{Fit83,ZloCamLou11,Donetal25}. The recoil velocity $v_{\rm esc}$ depends on a number of factors, including the ratio of the black hole masses, the values and orientations of their spins relative to the direction of the orbital angular momentum. The ejection velocities in the case of black holes with approximately equal masses range from hundreds to thousands of km per second. 

For the purposes of this article, the most interesting black holes are those ejected at velocities of the order of hundreds of km per second. We will now consider the part of the velocity distribution where black holes have such velocities. This speed is comparable to the characteristic velocity dispersion in the central region of the galaxy at a distance of several parsecs, i.e., the ejected SMBH does not leave the galaxy and may not even fly far from its center. The important thing is that at a distance of about 1--3~pc from the center, the SMBH's speed relative to the gas is several hundred km per second. At such distances, the gas density may already be several orders of magnitude lower than near the center, but it is still quite high. Under quite typical conditions at such distances, SMBHs actively collect gas through the process of ordinary Bondi accretion. 

A characteristic feature of SMBH ejection during the merger of a pair is the loss of the accretion disk, which is highly likely to have been destroyed during the approach and merger of SMBHs. The resulting SMBH is ejected without a disk. Furthermore, most of the surrounding star cluster was also dispersed during the SMBH's approach, and the resulting SMBH probably does not carry a significant number of stars with it when it is ejected. However, as a result of the relatively low ejection velocity and the accretion of surrounding gas, an accretion disk should form around the SMBH again, and the nucleus's activity should resume. Unlike the SMBH in the center, this process will occur at a lower density of the surrounding gas and outside the dust torus. As we will show later, these conditions are favourable for accelerating the UHECRs.

Let us make some general estimates. The mass growth due to Bondi accretion \cite{Bon52} has the rate
\begin{equation}
\frac{dM}{dt}=\frac{4\pi\rho G^2M^2}{(v_{\rm esc}^2+c_s^2)^{3/2}},
\label{acceq}
\end{equation}
where the sound speed is $c_s=[5k_{\rm B}T_i/(3m_p)]^{1/2}$, $k_{\rm B}$ is the Boltzmann constant, $\rho=nm_p$, $m_p$ is the mass of a proton, and $n$ is the number density of protons of the gas. For a gas with a temperature of $T=10^4$~K, the speed is $c_s\simeq12$~km~s$^{-1}$. In the case of $v_{\rm esc}\gg c_s$, we have
 \begin{equation}
\frac{dM}{dt}\simeq0.7\left(\frac{M}{10^8M_\odot}\right)^2\left(\frac{n}{10^2\mbox{~cm$^{-3}$}}\right)\left(\frac{v_{\rm esc}}{200\mbox{~km~s$^{-1}$}}\right)^{-3}\frac{M_\odot}{\mbox{yr}}.
\label{acceq2}
\end{equation} 
Thus, we can see that in the region a few parsecs from the center, there are conditions for sufficiently strong accretion and the resumption of activity of the ejected SMBH.


  \subsection{Slingshot ejection and the remnant of the star cluster}
  \label{rogatsec}
  
Let us now consider the case of a tidal ejection during a triple approach of SMBHs. In this case, a pair of SMBHs and a single SMBH are ejected from the center of the galaxy. Let us estimate the parameters of the star cluster remnant that the single SMBH carries with it. 

The observed parameters of the central star clusters in galaxies are described, for example, in \cite{Boe10}. Along with the extended part of the cluster with a radius of $\sim3-5$~pc, there is a stellar cusp in the vicinity of the SMBH, in which stars move in Keplerian orbits around the SMBH. If the average number density of stars in the region of the central parsec has a typical value of $n_*\sim10^7$~pc$^{-3}$, in the cusp it reaches $n_*\sim(10^8-10^9)$~pc$^{-3}$. Below, for example, we choose a fairly optimistic option with $n_*\sim10^8$~pc$^{-3}$.

Let us consider a simple model situation where all three SMBHs have the same masses $M$. Two of them formed a pair, and as they approached the third SMBH, one of the three SMBHs flew out during a slingshot and carried away its accretion disk and part of the stars from the central region of the star cluster. It was shown in \cite{LinSas77} that the probability of accretion disk survival during the ejection is high. In this paper, we will show that in addition to the disk, a part of the stellar cusp can survive and fly out along with the SMBH. As we will show later, this ejected part of the cluster probably does not participate in the reactivation of the ejected galactic nucleus; however, calculating the properties of the ejected cluster is interesting in its own right. 

During the slingshot ejection (in the case of triple approaches), the typical velocity of the ejected SMBH is $v_{\rm esc}\sim1000$~km~s$^{-1}$ and there is a wide distribution over $v_{\rm esc}$, see Fig.~11 in \cite{HofLoe07}. The pair of SMBHs responsible for the ejection should have approximately the same orbital velocity, and then the radius of its orbit is
\begin{equation}
r_{\rm esc}\sim \frac{GM}{4v_{\rm esc}^2}.
\end{equation}
Some of the stars within this radius may be lost after the slingshot, as they become unbound. The exact proportion of the remaining stars and the structure of the surviving part of the cluster is a complicated dynamic problem that can only be investigated using large numerical simulations. In this paper, we assume that a part of the cluster survives at a distance $r_{\rm cl}=f_{\rm cl}r_{\rm esc}$, where we leave the parameter $f_{\rm cl}$ free. Thus, the total radius of the remnant cluster, which was carried away along with the ejected SMBH, is equal to 
\begin{equation}
r_{\rm cl}\sim2\times10^{-2}\left(\frac{M}{10^8M_\odot}\right)\left(\frac{f_{\rm cl}}{0.2}\right)\left(\frac{v_{\rm esc}}{1000\mbox{~km~s$^{-1}$}}\right)^{-2}~\mbox{pc}.
\end{equation}
The typical size of the accretion disk does not exceed $\sim10^{-3}$~pc \cite{Moretal10}, so there may be a radius interval outside the accretion disk containing the remnant of the stellar cusp. 

The number of remaining stars in the cluster after the ejection is $N_*=4\pi r_{\rm cl}^3 n_*/3$, numerically
\begin{eqnarray}
N_*\simeq4\times10^3\left(\frac{M}{10^8M_\odot}\right)^3\left(\frac{f_{\rm cl}}{0.2}\right)^3
\nonumber
\\
\times\left(\frac{n_*}{10^8\mbox{~pc$^{-3}$}}\right)\left(\frac{v_{\rm esc}}{1000\mbox{~km~s$^{-1}$}}\right)^{-6}.
\label{nzv}
\end{eqnarray}
The characteristic velocity of stars in the cluster is $v_{\rm cl}\sim(GM/r_{\rm cl})^{1/2}$, the dynamic time in the cluster is $t_{\rm d}\sim r_{\rm cl}/v_{\rm cl}$, and the time of two-body dynamic relaxation in the case $N_*m\ll M$ can be written as 
\begin{equation}
t_{\rm r}\simeq\frac{M^2}{m^2}\frac{1}{\lambda_1N_*}t_{\rm d},
\label{trel}
\end{equation}
where $\lambda_1\simeq2\pi\ln[(4\pi)^{-1/2}(M/m)^{3/2}N_*^{-1/2}]$, and $m$ is a typical star mass. Note that the lifetime of a cluster is $t_{\rm ev}\simeq40t_{\rm r}$ in the so-called evaporative model, but in other models it reaches $\sim300t_{\rm r}$. Stars in finite orbits fall onto the SMBH due to their diffusion through the boundary of the loss cone. The rate of this diffusion is $\dot N_{\rm dif}\simeq N_*/(\lambda_2t_{\rm r})$ \cite{Dok91}, where $\lambda_2=\ln(r_{\rm cl}/r_{\rm t})$, and $r_{\rm t}$ is the radius of the tidal destruction of a star. According to \cite{Dok91}, $r_{\rm t}\simeq2r_{\rm s}(M/M_\odot)^{1/3}$, where $r_{\rm s}$ is the radius of the star. 

Unfortunately, the relaxation time calculated using the formula (\ref{trel}) is too long, so the rate $\dot N_{\rm dif}$ at which the accretion disk is supplied with stellar matter is very low. Resonant relaxation, if it occurs, reduces the relaxation time by one factor $M/m$, but this time still remains very long. The long relaxation time is due to the presence of a mass‑dominant black hole. Thus, the resumption of activity in the ejected nucleus is most likely not due to processes in the remnant stellar cluster, but to the accretion of gas located along the path of the SMBH, as described in the previous Subsection.


\section{Acceleration of cosmic rays at a shock wave}
\label{uhecrsec}  

Consider the UFO from the ejected galactic nucleus. We are interested in the process of cosmic ray acceleration at the shock wave of the outflow front in comparison with a similar process in the case of conventional AGN. Therefore, the calculation is done first for the conventional AGN, taking into account the presence of a dusty torus around it, and then a similar calculation is performed for the ejected core when it has moved away from the torus and entered a region with a significantly lower gas density. If the accretion disk is being replenished at this time and a UFO is initiated, then, as we will show, cosmic rays will be accelerated more efficiently than in the usual AGN. The activity of the ejected galactic nucleus will likely be intermittent, as the nucleus will pass through regions of low density and through denser clouds, which occupy $\sim10$~\% of the volume. Each passage episode may be accompanied by a new UFO and a new shock wave.

At the beginning of the active phase, a UFO occurs near the inner part of the accretion disk under the influence of radiation pressure and, possibly, magnetohydrodynamic processes. The gas flows out in the form of wide cones on both sides of the accretion disk, and a shock wave appears at the outflow front. For calculations, we will use the generalized Sedov self-similar solution for the shock wave. Sedov's classical solution assumed an explosion with instantaneous energy release, but in the work of V.I.~Dokuchaev \cite{Dok02} this solution was generalized for the case of continuous energy release. We will consider the Dokuchaev solution with constant energy release $\dot E=const$. In this case, the radius of the shock front depends on time as follows \cite{Dok02}
\begin{equation}
r(t)=a\left(\frac{\dot E t^3}{\rho}\right)^{1/5},
\label{doksol}
\end{equation}
where $a\simeq0.929$ \cite{Dok02}, $\rho=nm_p$ is the mass density of the gas in which the shock wave propagates. Next, we assume $a=1$ everywhere, which means a renormalization $\rho\to \rho a^{-5}$. The velocity of the shock wave in a typical situation is much lower than the speed of light and decreases with time according to the law $\propto t^{-2/5}$. As a concrete example, we assume that the bolometric luminosity is a fraction of $f_{\rm bol}\sim0.1$ of the Eddington luminosity, $L_{\rm bol}=f_{\rm bol}L_{\rm Edd}$, where $L_{\rm Edd}=1.26\times10^{46}(M/10^8M_\odot)$~erg~s$^{-1}$. The UFO power is $\dot E=f_{\rm kin}L_{\rm bol}$, where $f_{\rm kin}\sim0.05$. Let us also take the standard assumption of the mass flow onto a BH $\dot M=L_{\rm Edd}/(\eta c^2)$, where $\eta\simeq0.06$.

In the region of compression, the magnetic field increases. Following the usual assumptions of \cite{WanLoe17}, supported by observational data, we assume an approximate energy equipartition with a fraction of the energy of the magnetic field $\xi_B\sim0.1$  
\cite{WanLoe17}. If the mean UFO energy density is $\varepsilon\simeq3\dot E t/(4\pi r^3)$, then the magnetic field strength $B$ is found from the condition
\begin{equation}
\frac{B^2}{8\pi}\simeq \xi_B\varepsilon.
\label{magfield1}
\end{equation}
Numerically, we obtain
\begin{eqnarray}
B=\frac{6^{1/2}\xi_B^{1/2}\rho^{1/6}\dot E^{1/3}}{r^{2/3}}\simeq0.1\mbox{~G}\left(\frac{\xi_B}{0.1}\right)^{1/2}
\label{magfield2}
\\
\times\left(\frac{n}{10^6\mbox{~cm$^{-3}$}}\right)^{1/6}\left(\frac{\dot E}{10^{43}\mbox{~erg~s$^{-1}$}}\right)^{1/3}\left(\frac{r}{1\mbox{~pc}}\right)^{-2/3},
\nonumber
\end{eqnarray}
where we expressed $t$ in terms of $r$, according to (\ref{doksol}).

\begin{figure}
	\begin{center}
\includegraphics[angle=0,width=0.5\textwidth]{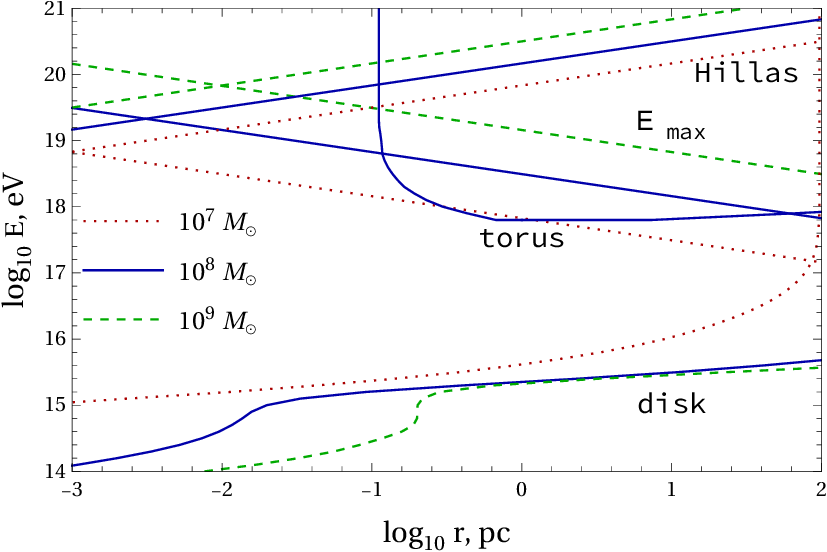}
	\end{center}
\caption{Constraints on the energy of accelerated protons from an ordinary active nucleus in the center of the galaxy, where the number density of the gas particles is $10^6$~cm$^{-3}$. The dashed, solid, and dotted lines correspond to the masses of the central black holes $M=10^9M_\odot$, $10^8M_\odot$, and $10^7M_\odot$. Decreasing straight lines show the maximum energy achievable when accelerated by an UFO shock wave. The increasing straight lines show the constraint according to the Hillas criterion. The upper limits on the radiation of the accretion disk are also shown. The bends on the curves correspond to the transition from the birth of $e^+e^-$ to photohadronic processes. For the case of $M=10^8M_\odot$, an example of a constraint on the radiation of a torus is shown.}
	\label{grcon1}
\end{figure}

\begin{figure}
	\begin{center}
\includegraphics[angle=0,width=0.5\textwidth]{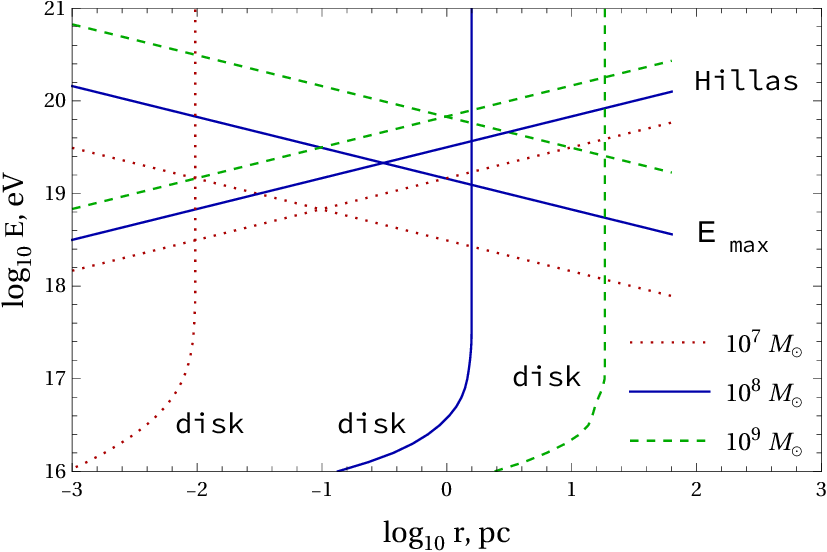}
	\end{center}
\caption{The same but for an ejected nucleus moving in a gaseous medium with a particle number density of $10^2$~cm$^{-3}$. The distance $r(t)$ denotes the current radius of the shock wave, according to the Dokuchaev solution (\ref{doksol}).}
	\label{grcon2}
\end{figure}

Following the work of \cite{WanLoe17}, we assume that proton acceleration occurs via the first-order Fermi mechanism with the Bohm diffusion coefficient. For simplicity, in this article we consider only the acceleration of protons, but not heavy nuclei. The maximum energy of a proton at the moment $t$ is
\begin{equation}
E_{\rm max}=e Bv_{\rm s}^2t/c,
\label{emax1}
\end{equation}
where $v_{\rm s}=dr(t)/dt$ is the velocity of the shock wave according to (\ref{doksol}). Substituting the shock wave parameters, we obtain
\begin{equation}
E_{\rm max}=\frac{2^{1/2}3^{5/2}}{5^{1/2}}\frac{e}{c}\xi_B^{1/2}\rho^{-1/6}\dot E^{2/3}r^{-1/3}.
\label{emax2}
\end{equation}
This value is shown in Fig.~\ref{grcon1} and Fig.~\ref{grcon2} for different values of $\rho$ at different distances from the center. We emphasize that the values in Fig.~\ref{grcon1} and Fig.~\ref{grcon2} are shown for the conditions at the front of a shock wave moving according to the law (\ref{doksol}); therefore, different times $t$ correspond to the same distances $r$ in two cases, which is crucial for the acceleration of cosmic rays, since the characteristic acceleration time is compared with the characteristic loss time and with the dynamic time.  It can be seen from the Figures that the maximum energy decreases with distance, and the highest energies of accelerated protons could be achieved at short distances. However, the energy gain to the value (\ref{emax2}) is hindered by a number of circumstances, which we will now consider.

First of all, there is a classical energy limitation -- the Hillas criterion (the size of the acceleration region must exceed the gyroradius) $E_{\rm max}<eBr$. The Hillas criterion constraint is also shown in Fig.~\ref{grcon1} and Fig.~\ref{grcon2}. Comparing the Hillas criterion with the expression (\ref{emax2}) gives a universal value of the constraint, independent of $\rho$, in the form
\begin{eqnarray}
&&E_{\rm max}=\frac{2^{1/2}3^{3/2}}{5^{1/4}}\frac{e}{c^{1/2}}\xi_B^{1/2}\dot E^{1/2}
\label{emaxhill}
\\
&\simeq&2.1\times10^{19}\left(\frac{\xi_B}{0.1}\right)^{1/2}\left(\frac{\dot E}{6.3\times10^{43}\mbox{~erg~s$^{-1}$}}\right)^{1/2}\mbox{~eV}.
\nonumber
\end{eqnarray}
We emphasize that this restriction applies only to acceleration in the shock wave in the form of (\ref{doksol}) and is applicable only at the front of the shock wave. The constraint (\ref{emaxhill}) corresponds to the specific radius
\begin{equation}
r=\frac{3^3}{5^{3/4}c^{3/2}}\rho^{-1/2}\dot E^{1/2},
\end{equation}
which corresponds to the intersection points of the straight lines in Fig.~\ref{grcon1} and Fig.~\ref{grcon2}.

The energy losses of ultrahigh-energy protons occur due to the generation of electron-positron pairs and, at high energies, due to pion photoproduction. In \cite{BerGazGri06}, the inverse energy loss time of protons with energy $E$ in microwave background radiation with temperature $T_{\gamma 0}=2.7$~K was calculated as
\begin{equation}
\beta_0(E)=-\frac{1}{E}\frac{dE}{dt},
\label{beta0}
\end{equation}
and a recipe for recalculating losses is given for the case of proton motion in a blackbody radiation field with a temperature of $T$
\begin{equation}
\beta(E,T)=\left(\frac{T}{T_{\gamma 0}}\right)^3\beta_0\left[\left(\frac{T}{T_{\gamma 0}}\right)E\right].
\label{betat}
\end{equation}
Figure~1 of \cite{BerGazGri06} shows the function (\ref{beta0}), taking into account both the birth of $e^+e^-$ pairs and photohadronic losses. We will use this recipe to calculate proton energy losses due to radiation from an accretion disk with inner and outer radii $R_{\rm min}$ and $R_{\rm max}$, respectively. The inner radius is determined by the last circular orbit $R_{\rm min}=3R_g=6GM/c^2$. Consider a proton at a distance of $r\gg R_{\rm max}$ from the center of the disk. Each annular sector of the disk with a radius of $R$ produces a photon flux with some effective temperature of $T(R)$, so that when the loss rates are added together, we get the inverse of the radiation loss time of the entire disk.
\begin{equation}
\frac{1}{t_{\rm disk}(E)}
=\frac{1}{4\pi r^2}\int\limits_{R_{\rm min}}^{R_{\rm max}}2\pi RdR\beta(E,T(R)),
\label{tdisk}
\end{equation}
where $T(R)=[Q(R)/\sigma]^{1/4}$, $\sigma$ is the Stefan--Boltzmann constant, and the rate of energy release per unit surface of the disk is \cite{ShaSun73}
\begin{equation}
Q(R)=\frac{3GM\dot M}{8\pi R^3}\left[1-\left(\frac{R_{\rm min}}{R}\right)^{1/2}\right],
\label{qeq}
\end{equation}
where $\dot M$ is the accretion mass flow, and the function $\beta$ from \cite{BerGazGri06} is recalculated for the temperature of each annular sector. The constraints on $E_{\rm max}$ due to disk radiation play a major role. They are shown in Fig.~\ref{grcon1} and Fig.~\ref{grcon2}, and then we will discuss them in more detail.

There is also an extended dust torus in the central region of the AGN, which re-emits the hard radiation of the accretion disk in the infrared range. If the torus has a characteristic size of $R_{\rm tor}$, then for $r\leq R_{\rm tor}$ it can be assumed that the radiation flux is approximately constant and corresponds to blackbody radiation with a characteristic temperature $T_{\rm tor}\sim300$~K. In this case, the inverse characteristic loss time at the torus for particles with energy $E$ can be estimated as 
\begin{equation}
\frac{1}{t_{\rm tor}}\simeq\beta(T_{\rm tor},E).
\end{equation}
But when the ejected core flies out to a distance of $r\gg R_{\rm tor}$, the photon flux from the torus will decrease as $\sim R_{\rm tor}^2/r^2$. Taking these effects into account, it is possible to calculate the radiation limit of the torus depending on the distance.  Fig.~\ref{grcon1} shows an example of such a constraint for $M=10^8M_\odot$. In the case of an ejected core that flew out at a distance of $\gg R_{\rm tor}$, there is practically no such constraint; therefore, in Fig.~\ref{grcon2} it is not shown. It can be concluded that the radiation limit of the torus is weaker than the radiation limit of the disk. For the case of an ordinary galactic nucleus, even the torus constraint would limit the maximum energy of accelerated protons (although the restriction on disk radiation is more significant), whereas in the case of ejected nuclei there is no torus constraint. In addition, radiation from the disk's corona and pp collisions contribute to the constraints, which turn out to be insignificant. Nevertheless, $pp$ collisions could lead to the generation of neutrino fluxes; therefore, in the mechanism we are considering, neutrino fluxes are reduced, however, a detailed consideration of neutrino generation is beyond the scope of this work.

There is a constraint on the disk, but it allows the acceleration of protons to energies about an order of magnitude higher than in ordinary (non-ejected) galactic nuclei. The physical reason for this is that the ejected core falls into an area with a lower gas density. At a lower density, the velocity of the shock wave is higher (see (\ref{doksol})). This leads to the fact that the shock wave quickly moves away from the accretion disk, the effect of the hard radiation of the disk lasts less time, and therefore the protons have time to accelerate to a higher energy. 

\begin{figure}
	\begin{center}
\includegraphics[angle=0,width=0.5\textwidth]{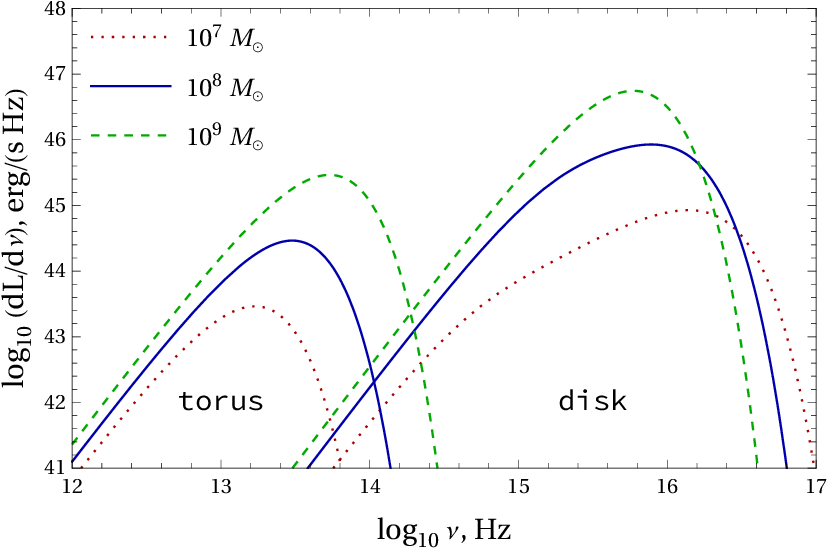}
	\end{center}
\caption{Radiation spectrum of an accretion disk and a torus in the nucleus of an active galaxy. The dashed, solid, and dotted lines correspond to the masses of the central black holes $M=10^9M_\odot$, $10^8M_\odot$, and $10^7M_\odot$. The accretion parameters are listed in the text of the article.}
	\label{grspectr}
\end{figure}

By integrating over the annular sectors of the accretion disk and over the characteristic volume of the torus, it is possible to obtain the radiation spectrum of the disk and the torus. This spectrum is shown in Fig.~\ref{grspectr} depending on the mass of the central black hole. The calculated spectrum coincides qualitatively with that given in \cite{EhlOikPer25}.
   
According to Fig.~\ref{grcon1} and Fig.~\ref{grcon2}, it is possible to compare the maximum energies of accelerated protons in the case of an ordinary nucleus of an active galaxy and an ejected nucleus. 
The speed of shock wave propagation in these two cases depends on the number density of atoms in the galaxy. Approximately, this density can be represented in the following form \cite{MatBak71,MatLoe86,ValBas91}
\begin{equation}
n\simeq10^{-2}\left(\frac{r}{1\mbox{~kpc}}\right)^{-2}\mbox{~cm$^{-3}$}.
\end{equation}
As can be seen from the figures, the constraint on the radiation of the disk in an ordinary nucleus significantly limits the energy of the proton. Only at a distance of $\sim10^2$~pc from the center would acceleration to an energy of $\sim10^{17}$~eV be possible. In this case, a more accurate calculation gives energy up to $\sim10^{18}$~eV \cite{Peretal23}, \cite{EhlOikPer25}. On the contrary, in the case of the ejected nucleus shown in Fig.~\ref{grcon2}, already at a distance of $\sim2-3$~pc for $M=10^8M_\odot$ the energy $\sim10^{19}$~eV is achieved, and the effect of acceleration stochasticity, similar to that considered in \cite{Peretal23}, \cite{EhlOikPer25}, can give even greater energies.


\section{Conclusion}

The UHECRs acceleration in a spherical shock wave created by the UFO from the vicinity of SMBHs in the centers of AGNs was discussed in \cite{WanLoe17,Peretal23,EhlOikPer25}. The maximum energy of the accelerated particles in this model is limited by the processes of their interaction with the strong radiation field produced by the accretion disk, the corona of the disk, and the dusty torus. Whereas \cite{WanLoe17} optimistically showed that protons can be accelerated to energies $\sim10^{20}$~eV, \cite{Peretal23}, \cite{EhlOikPer25} demonstrated acceleration only to $\sim10^{18}$~eV by numerical modeling, and only a small fraction of protons manage to achieve higher energies by diffusion.

In this paper, we point out that there is a subpopulation of active galactic nuclei in which protons are capable of accelerating to energies of $\geq10^{19}$~eV. This subpopulation represents the cores of galaxies ejected from the center during binary and triple approaches of SMBHs. The  approaches and mergers of SMBHs are  a typical and fairly common process in the hierarchical picture of galaxy formation. In a slingshot, a black hole can take with it the accretion disk and part of the central star cluster, but the dusty torus is located outside the tidal radius and therefore does not follow the SMBH. The accretion of surrounding gas onto the ejected SMBH can reactivate it if the SMBH is moving in an area with a sufficiently high gas density and at a not‑too‑high speed.

Some time after the ejection, the ejected active nuclei fall into an area with a lower gas density, where the velocity of the shock wave is higher. If the ejected nucleus enters the active phase after moving a considerable distance from the center, the resulting shock wave propagates at high speed, moving away from the accretion disk. At the same time, the time of interaction of protons with the intense radiation of the disk decreases, and protons leave the area of strong influence of the disk faster. This creates conditions for accelerating protons to high energies, so the ejected nuclei can be effective accelerators of UHECRs. Thus, the main advantage of the ejected AGNs as UHECR accelerators is the low density of the medium and the absence of a dusty torus, the radiation of which also limits the maximum energy of the accelerated particles at a distance of about a parsec from the black hole. This solves the problem of limiting the maximum particle energy found in the \cite{WanLoe17,EhlOikPer25} models, where the shock wave remains in the area of the disk radiation for a long time.

From the point of view of observations, there are systems whose properties can be explained by the slingshot ejection of galactic nuclei, although it is not yet possible to draw a final conclusion. In \cite{ValBas91}, a binary quasar is discussed, the components of which are very close in angular distance, but have different redshifts. It is assumed that one of these components is a galactic core ejected at a relativistic speed. In the works \cite{SasValAar74,LinSas77,ValKot89,MikVal90}, ejected SMBHs are considered as energy sources for some radio galaxies with a binary structure. An ejected black hole with a mass of $\sim10^5M_\odot$ was identified in the work of \cite{Linetal16}. The presence of X-ray radiation indicates that the accretion-related material has been preserved around the black hole. There is also observational evidence of the presence of three SMBHs in a distant galaxy at the redshift $z\sim5$ \cite{Ubletal26}. However, the number of such objects is poorly known, and reliable conclusions cannot yet be drawn about the statistics of UHECRs accelerated in these objects. 

Recent observations by the James Webb Space Telescope (JWST) have yielded new and interesting data that have not yet been fully explained. As predicted by the theory of galaxy formation and mergers, 1-5~\% of AGNs should be binary, located at a distance of less than 30~kpc from each other. JWST observations have revealed four double AGNs, and a triple nucleus was detected using a combination of data from JWST and the VLT telescope \cite{Peretal25}. Taking into account the observed number and the probability of AGN detection, it was concluded that the number of multiple AGNs in the early Universe is $\simeq20-30$~\%, significantly higher than the expected value. It is still unknown what processes could have led to this number.

Although the ejected nuclei are rare objects, each ejection almost certainly triggers the activity. Therefore, the number of active ejected nuclei can be comparable to the number of active stationary ones. Nevertheless, the calculation of statistics on such events requires further research. In this paper, we have only shown that when accelerated at the UFO front from ejected galactic nuclei, protons can gain about an order of magnitude more energy than in the original galactic nuclei.


\end{document}